\documentclass[journal]{IEEEtran}
\usepackage{xcolor,soul,framed} 
\usepackage{cite}
\usepackage{hyperref}
\colorlet{shadecolor}{yellow}
\usepackage[pdftex]{graphicx}
\usepackage{float}
\graphicspath{{../pdf/}{../jpeg/}}
\DeclareGraphicsExtensions{.pdf,.jpeg,.png}
\usepackage[ruled,vlined]{algorithm2e}
\usepackage{balance}
\usepackage{array}
\usepackage{tikz}
\usetikzlibrary{shapes,arrows}
\tikzstyle{decision} = [diamond, draw, fill=blue!20, 
text width=4.5em, text badly centered, node distance=3cm, inner sep=0pt]
\tikzstyle{block} = [rectangle, draw, fill=blue!20, 
text width=5em, text centered, rounded corners, minimum height=4em]
\tikzstyle{line} = [draw, -latex']
\tikzstyle{cloud} = [draw, ellipse,fill=red!20, node distance=3cm,
minimum height=2em]
\usepackage[cmex10]{amsmath}
\usepackage{array}
\usepackage{mdwmath}
\usepackage{mdwtab}
\usepackage{eqparbox}
\usepackage{url}
\usepackage{subcaption}
\title{Detection of Seismic Infrasonic  Elephant Rumbles Using Spectrogram-Based Machine Learning}
\author{A.~M.~J.~V.~Costa, C.~S.~Pallikkonda, H.~H.~R.~Hiroshan, G.~R.~U.~Y.~Gamlath, S.~R.~Munasinghe,~\IEEEmembership{SMIEEE}, C.~U.~S.~Edussooriya~\IEEEmembership{}
\thanks{Manuscript received mm dd, yyyy. All authors are affiliated to the Department of Electronic and Telecommunication Engineering, University of Moratuwa, Moratuwa 10400, Sri Lanka}
\thanks{S.~R.~Munasinghe is a visiting Fellow at the Department of Global Development, CALS, Cornell University, Ithaca 14850, NY, USA (e-mail: srm278@cornell.edu)}}
\begin{document}
\maketitle
\begin{abstract}
This paper presents an effective method of identifying elephant rumbles in infrasonic seismic signals. The design and implementation of electronic circuitry to amplify, filter, and digitize the seismic signals captured through geophones are presented. A collection of seismic infrasonic elephant rumbles was collected at a free-ranging area of an elephant orphanage in Sri Lanka. The seismic rumbles were converted to spectrograms and several methods were used for spectral feature extraction. Using {\tt LasyPredict}, the features extracted using different methods were fed into their corresponding machine-learning algorithms to train them for automatic seismic rumble identification. It was found that the Mel frequency cepstral coefficient (MFCC) feature extraction method and the Ridge classifier machine learning algorithm give the best performance in identifying seismic elephant rumbles. A novel method for denoising the spectrum that leads to enhanced accuracy in identifying seismic rumbles is also presented.
\end{abstract}
\begin{IEEEkeywords}
	Infrasonic elephant rumbles, spectrogram, CMRR,  MFCCs, Hjorth parameters, Geophone, seismic waves 
\end{IEEEkeywords}
\IEEEpeerreviewmaketitle
\section{Introduction}
\IEEEPARstart{E}lephants use different vocalization patterns to communicate between herds and each other for food, water, mating, and warning. Among these patterns, the infrasonic rumbles have a very specific significance in long-range communication \cite{nair2009vocalizations}. Elephant rumbles have acoustic and seismic components. The acoustic component is what propagates through the air as a 3D wave, and gets attenuated through the foliage fairly quickly. The seismic component on the other hand propagates through the ground as a 2D Rayleigh wave and therefore, travels a longer distance compared to the acoustic component. Elephant's foot is believed to have the capability of a seismic transponder to facilitate long-range seismic infrasonic communication. These seismic signals can be a powerful component of an elephant’s communication system, serving crucial functions such as mate finding, prey detection, and interspecific and intraspecific warnings \cite{o2000seismic}. The attenuation of seismic waves during transmission increases monotonically with frequency, hence the range of 10Hz to 40Hz is considered the "sweet zone" for seismic signal propagation. Elephant rumbles, which have a fundamental frequency of around 20Hz \cite{payne1986infrasonic}, fall within this sweet zone, allowing them to communicate through the ground. The elephant rumbles that propagate as Rayleigh waves through the ground can be captured using geophones \cite{gunther2004seismic}.\par
Detecting and localizing wild elephants through airborne rumbles encounter certain challenges that impede its effectiveness. Airborne signals, though easy to detect, attenuate rapidly limiting the range over which elephants can be identified. The amplitude of a Rayleigh wave during ground surface transmission is inversely proportional to
the square root of the distance, with a loss of 3dB for every doubling of the distance. In airborne signals,
amplitude is inversely proportional to the distance, with a loss of 6dB for every doubling of the distance. Factors such as weather, time of day, scattering, and reflections due to objects, and trees also attenuate the airborne signal \cite{o2000seismic}. The seismic wave undergoes a lower attenuation compared to the airborne wave, hence the detectable range of the seismic component of a rumble is greater than that of the airborne acoustic component \cite{sayakkara2017eloc}. A very few prior researches have demonstrated the potential of using infrasonic seismic rumbles in detecting the presence of wild elephants \cite{zeppelzauer2015towards}, \cite{Reinwald2021}. In this context, this research aims at developing technology for detecting infrasonic elephant rumbles with the expectation of devising a tool for the localization of wild elephants, hence mitigating the raging human-elephant conflict in some parts of the world.\par
Seismic waves of foot stomps can also be considered for elephant detection, but foot stomping being a short-duration event, its seismic signal undergoes a greater dispersion which leads to widening the signal bandwidth compared to that of the rumble. This paper presents the electronic circuitry and signal processing involved in dealing with infrasonic seismic signals. A novel method to enhance the seismic spectrogram, hence to improve rumble identification ability is also presented.
\section{Spectrogram of Infrasonic Seismic Signals}
\subsection{Overview of Signal Acquisition and Processing}
The seismic elephant rumbles are in the micro-volte range. They are picked up by geophones, and amplified electronically to a level suitable for signal processing. The amplified seismic signal is then bandpass filtered to within the 5Hz to 150Hz range, which is the typical bandwidth of elephant infrasonic rumble. This filtering ensures that noise and other unwanted signals are removed, enhancing the signal quality. To achieve the desired voltage gain, the signal is guided through a variable gain amplifier in which the gain is optimally adjusted. Then, the signal is sent to an analog-to-digital converter (ADC) for digitization. As an additional precaution, a voltage clipping stage is employed to safeguard against any potential voltage spikes that could adversely impact the data integrity. The digital signal is transmitted to a Raspberry Pi single-board computer through the I2C communication protocol. In the Raspberry Pi, the spectrogram features are extracted and fed to a trained machine-learned algorithm to determine if the spectrogram contains an elephant rumble.
\begin{figure*}
\centering
\includegraphics[width=1.8\columnwidth]{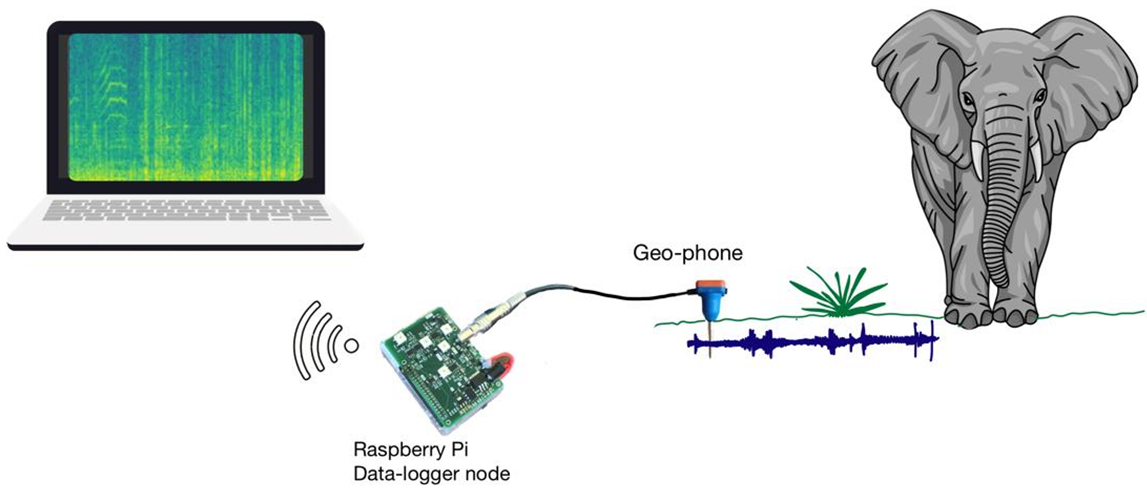} 
\caption{An overview of the elephant rumble detection system} \label{fig:dh-prodecrrnn}
\end{figure*}
\subsection {Electronic Circuitry}
\begin{figure}[H]
	\begin{center}
		\includegraphics[width=2.5in]{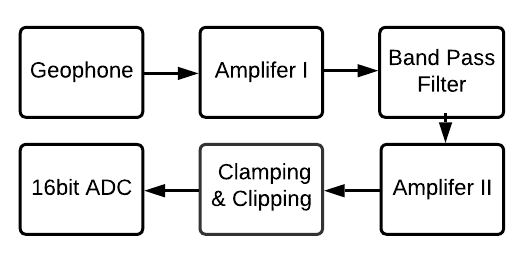}\\
		\includegraphics[width=3.2in]{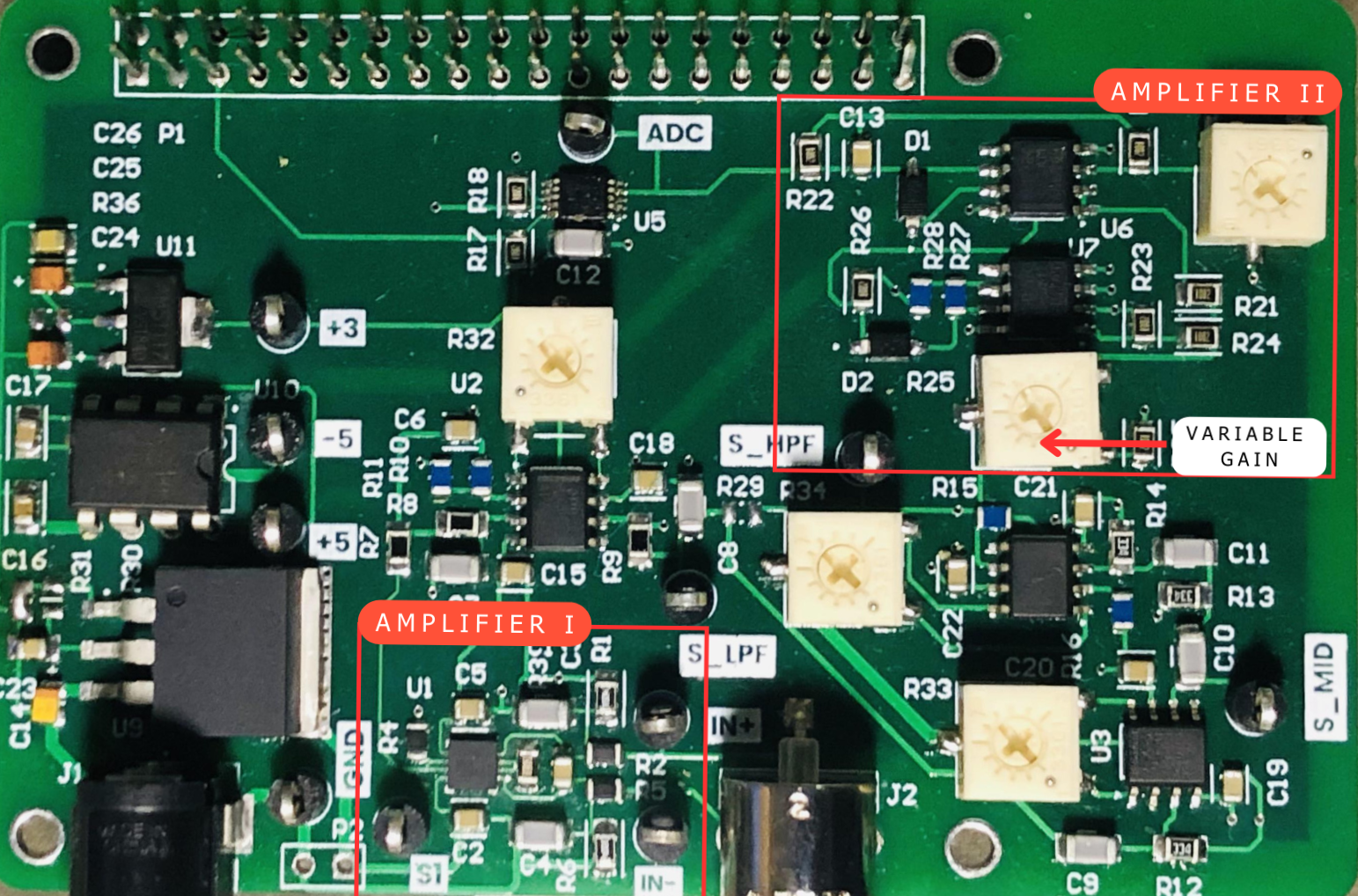}
		\caption{Seismic signal processing workflow (above), and the electronic circuit (below)}
		\label{hw design}
	\end{center}
\end{figure}

The geophone sensor is sensitive to the vertical component of the seismic signal (ground vibration). The lower microvolt analog signal it picks up varies in magnitude and frequency over a wide range, hence filtering and amplification are needed. Figure \ref{hw design} shows the seismic signal processing workflow. 
\subsubsection{Geophone Sensor}
A Geophone sensor consists of a coil of wire suspended between magnets with a mass attached to it. When seismic waves pass through the ground, the mass vibrates, generating electrical signals proportional to the vertical component of the ground vibration. In this research, SG-05 geophone was used. Technical specifications of the SG-05 geophone are listed in Table \ref{table:geophone}. 
\begin{table}[H]
	\caption{Geophone SG-05 Technical Specifications}
	\centering
	\begin{tabular}{l r}
		\hline\hline
		Property & Value\\
		\hline
		Natural frequency &  5 Hz\\
		Sensitivity &  80V/m/s\\
		Spurious frequency &  150 Hz\\
		Weight  &  22.7g\\
		Operating temprature  &  $-40^0c$ to $+80^0c$\\
		\hline
	\end{tabular}
	\label{table:geophone}
\end{table}
\subsubsection{Amplifier I}
The microvolt signal generated by the geophone is amplified by Amplifier I, which is an instrumental amplifier. It accepts a differential input and eliminates the common mode noise significantly with a high common mode rejection ratio(CMRR). Being an integrated circuit, it has a lower instrumental noise and offset. The passive components of the circuit were carefully selected considering the tolerances, as they can affect the signal quality significantly. The amplifier has a gain of 500 in this stage.
\subsubsection{Band Pass Filter}
\begin{figure}[H]
	\centering   \includegraphics[width=1\columnwidth]{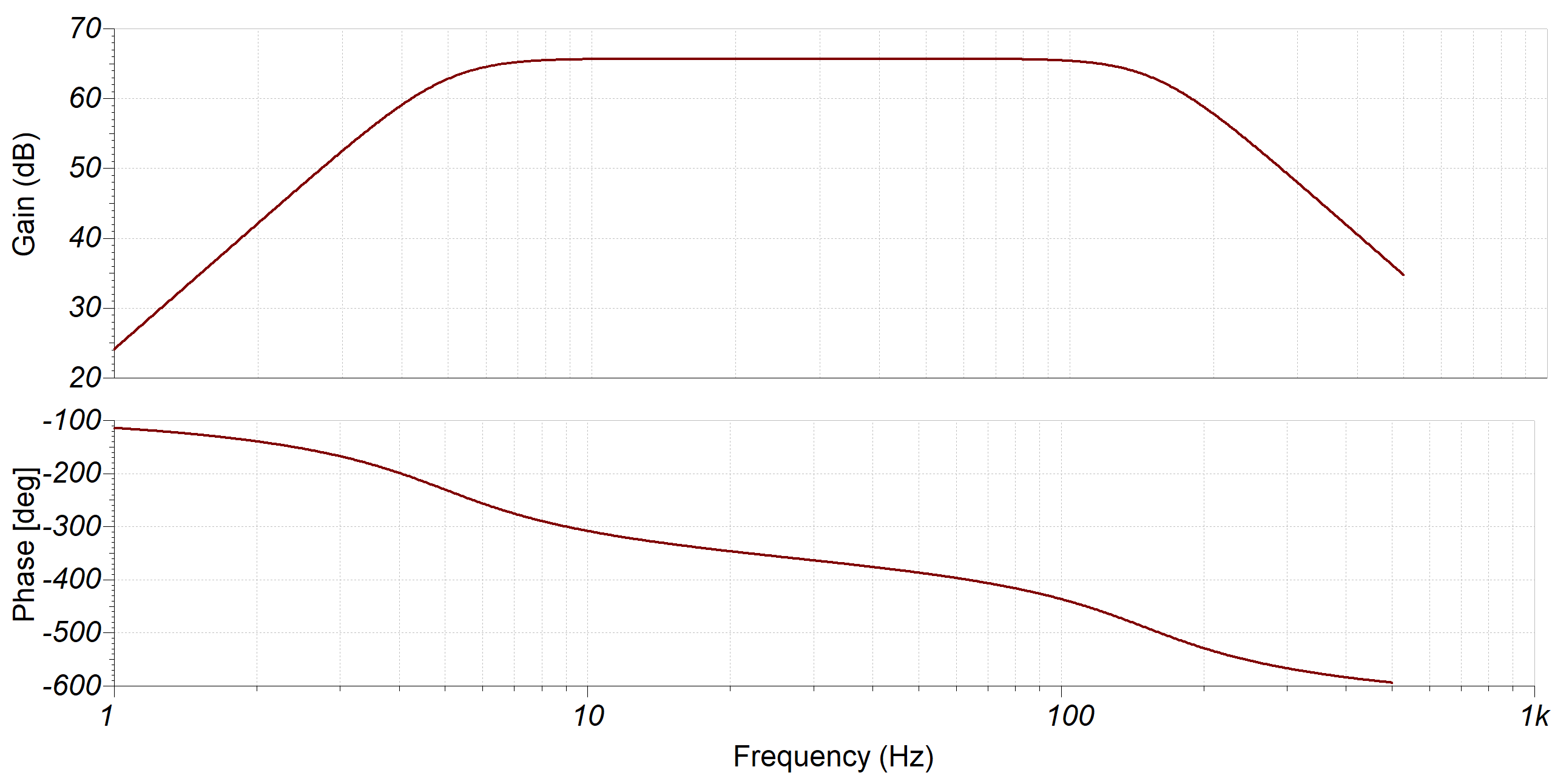}
	\caption{Frequency response of the Bandpass filter: magnitude(above) and phase(below) responses}
	\label{fig:bandpass1}
\end{figure}

The seismic elephant rumble spread over infrasonic frequencies, hence other frequencies the geophone picks up are to be removed. For this, a third-order Butterworth filter is used. Since the Butterworth filter has a flat passband gain, the filtering will uniformly affect the amplitudes of the passband frequencies in the geophone signal. The filter was designed in Sallen-key filter topology, using a low-noise operational amplifier (OPAmp)\cite{zumbahlen2007phase}. Figure \ref{fig:bandpass1} shows the gain and phase response of the filter 
\subsubsection{Amplifier II}
After the band-pass filtering, the signal is mostly infrasonic, and it may or may not contain an elephant rumble. The signal is still weak and therefore it is amplified again using the Amplifier II, which is an inverting amplifier with a variable gain that is adjusted using a variable resistor shown in Fig.\ref{hw design}(below). Amplifier II elevates the signal to the full voltage range of 0V–3.3V. The total amplification therefore is in the range of 3000-6000.
\subsubsection{Clamping and Clipping Circuit}
The voltage clipping circuit is designed to limit the analog signal to a safe range of 0V to 3.3V before feeding it into the analog-to-digital converter (ADC). This protective measure serves two primary purposes such as limiting the analog signal range and protecting against voltage spikes. Since the input of the ADC should be a positive voltage, the voltage clamp circuit is added to introduce a DC offset to the signal.
\subsubsection{Analog to Digital Converter (ADC)}
The 16-bit analog-to-digital converter (ADC) converts the amplified band-limited analog signal to digital format. Considering the signal bandwidth of 150 Hz, the sampling rate is selected as 475 samples per second so that aliasing is avoided. The digital data, in a 10Hz-150Hz band, is then transferred to the Raspberry Pi single-board computer for further processing. 
\subsubsection{Sensitivity}
The sensitivity of the 16-bit ADC with a range of 3.3 V can be obtained as
\begin{equation}
	S_{ADC}=\frac{3.3 \text{ V}}{2^{16} - 1} = 50.354\ \mu \text{V}\label{eq:Sadc}
\end{equation}
Given the geophone sensitivity $S_{G}$ and the amplification factor $g$, the sensitivity  of the system $S_{s}$ is given by 
\begin{equation}
	S_{s} = \frac{S_{ADC}}{gS_{G}}.
	\label{eq:sens}
\end{equation}
The sensitivity of the geophone is $S_{G}=80 \text{ V/ms}^{-1}$ (see Table \ref{table:geophone}), and with $g$=3000, $S_{s}=0.3147$ ms$^{-1}$. 
\subsection {Signal Processing}
\subsubsection{Wavelet Decomposition and Reconstruction}
The wavelet transform (WT) is used for the time-frequency decomposition of the seismic signal while also denoising the signal. The WT decomposes the signal into a sum of wavelets with varying positions and scales (frequencies), enabling to capture of low-frequency and high-frequency details. In this research, the Daubechies 3 wavelet \cite{Daubechies} wavelet decomposition is used.
\subsubsection{Identification of Frequency Ridges}
To identify and enhance ridge-like structures within the spectrogram, a ridge filter is first applied to identify ridge structures of the spectrogram. Then, the filtered spectrogram and the original spectrogram are combined using multiplicative blending. At each spatial position $(t, f)$ in the spectrogram $S(t, f)$, where $t$ represents time and $f$ denotes frequency, the partial derivatives (gradients) of the spectrogram were calculated with respect to both time and frequency. Then, the gradients are subjected to a two-dimensional Gaussian smoothing process. The structure tensor, which captures the local intensity distribution structure in the spectrogram's neighborhood is then constructed from the smoothed gradients as follows~\cite{zeppelzauer2015towards}:
\begin{align}
	T(t, f) =
	\begin{bmatrix}
		\nabla_t^2 & \nabla_t \cdot \nabla_f  \\
		\nabla_t \cdot \nabla_f  & \nabla_f^2
	\end{bmatrix}
\end{align}
where $T(t, f)$ is the structure tensor at $(t,f)$, $\nabla_t$  is the partial derivative with respect to time, and $\nabla_f$ is the partial derivate with respect to frequency. From the structure tensor, we derive the eigenvalues $\lambda_1$ and $\lambda_2$ using the following expression.
\begin{eqnarray}
	\lambda_{1,2}=\frac{1}{2}\left[\nabla_t^2 +\nabla_f^2 \pm \sqrt{(\nabla_t^2-\nabla_f^2)^2 + 4(\nabla_t \cdot \nabla_f)^2}\right]
\end{eqnarray}
When there is an edge-like structure such as a frequency contour in the local neighborhood of the spectrogram, the condition $\lambda_1$  $ >\lambda_2$ is fulfilled. In the case of a perfect edge, $\lambda_2$ becomes zero while the condition $\lambda_1$  $ >\lambda_2$  is still fulfilled. If  $\lambda_1$  $ =\lambda_2$, it indicates that the underlying structure is rotationally symmetric, such as an isolated spectral peak. If both eigenvalues become zero, it suggests that the underlying structure is homogeneous throughout the local neighborhood.
From the eigenvalues of the structure tensor at a given position $(t,f)$ in the spectrogram, the coherence is computed as follows:
\begin{equation}
	c = \frac{\lambda_1-\lambda_2}{\lambda_1+\lambda_2} 
\end{equation}
\subsubsection{Spectrogram Enhancement}\label{subsub-SE}
To enhance the visibility of frequency contours while simultaneously reducing noise, the coherence is used as a weighting function. The enhanced spectrogram, denoted as $\hat{S}(t, f)$, is computed as follows: 
\begin{equation}
	\hat{S}(t, f) = S(t, f) \cdot (c(t, f) + 1)   
\end{equation}
\begin{algorithm}[tb]
	\SetAlgoNlRelativeSize{-1}
	\caption{Enhanced Spectrogram Calculation}
	\KwData{\textit{s} - Input spectrogram\;}
	\KwResult{Enhanced spectrogram\;}
	Apply ridge filtering (\textit{Ridge\_s}) for \textit{s}\;
	Calculate time gradient (\textit{time\_gradient}) along axis 0\;
	Calculate frequency gradient (\textit{frequency\_gradient}) along axis 1\;
	Calculate tensor elements:
	\begin{align*}
		\textit{tensor\_xx} &= \textit{time\_gradient}^2 \\
		\textit{tensor\_yy} &= \textit{frequency\_gradient}^2 \\
		\textit{tensor\_xy} &= \textit{time\_gradient*frequency\_gradient} \\
	\end{align*}
	Apply Gaussian filtering to \textit{tensor\_xx}, \textit{tensor\_yy}, and \textit{tensor\_xy} to obtain smoothed versions\;
	Calculate eigenvalues $\lambda_1$ and $\lambda_2$ \; 
	Calculate coherence ($\textit{c}$) as:
	\[\textit{c} = \frac{\lambda_1 - \lambda_2}{\lambda_1 + \lambda_2}\]
	Calculate enhanced spectrogram ($\textit{enhanced\_spectrogram}$) as:
	\[\textit{enhanced\_spectrogram} = (10 \cdot \log_{10}(\textit{Ridge\_s})) \cdot (1 + \textit{c})\]
	\Return{\textit{enhanced\_spectrogram}}\;
\end{algorithm}

\begin{algorithm}[tb]
	\SetAlgoNlRelativeSize{-1}
	\caption{Threshold-Based Spectrogram Enhancement}
	\KwData{%
		\textit{s1} - Input spectrogram\;
		
	}
	\KwResult{%
		\textit{final\_spectrogram} - Further Enhanced spectrogram\;
	}
	
	Calculate Threshold values from \textit{s1} - \textit{threshold1, threshold2, threshold3} \;
	\For{$i$ in range($\textit{s1.shape[0]}$)}{
		\For{$j$ in range($\textit{s1.shape[1]}$)}{
			\If{$\textit{s1}[i, j] > \textit{threshold3}$}{
				$\textit{final\_spectrogram}[i, j] = \textit{s1}[i, j] + 5$\;
			}\ElseIf{$\textit{s1}[i, j] > \textit{threshold2}$}{
				$\textit{final\_spectrogram}[i, j] = \textit{s1}[i, j] + 2$\;
			}\ElseIf{$\textit{s1}[i, j] > \textit{threshold1}$}{
				$\textit{final\_spectrogram}[i, j] = \textit{s1}[i, j] - 2$\;
			}\Else{
				$\textit{final\_spectrogram}[i, j] = \textit{s1}[i, j] - 5$\;
			}
		}
	}
	\Return{\textit{final\_spectrogram}}\;
\end{algorithm}
Subsequently, three distinct percentiles ($25^{th}$, $50^{th}$, and $75^{th}$) are computed based on the input spectrogram and are used as threshold values. These percentiles effectively partition the spectrogram's pixel values into three categories: low, medium, and high intensity.
The algorithm iterates through each pixel within the spectrogram and applies adjustments contingent on the intensity of the pixel. Pixels with values surpassing the $75^{th}$ percentile undergo enhancement by adding 5, thus accentuating the regions with the highest intensity. Pixels falling within the range of the $50^{th}$ and $75^{th}$ percentiles experience a moderate enhancement of 2, ensuring that mid-level features remain discernible. Pixels below the $50^{th}$ percentile have their intensity reduced by subtracting 2, consequently amplifying the contrast between the background and the features of interest. Finally, for pixels with intensities below $25^{th}$ percentile, a reduction of 5 is applied, maintaining a clear distinction between the lowest-intensity areas and other regions of the spectrogram.
Additionally, following the enhancement process, a Gaussian blur operation is applied. This operation serves to further refine the spectrogram, smoothing out noise and improving its overall quality for more precise feature extraction and interpretation.
\subsubsection{Assessment of Spectrogram Enhancement}
To assess the effectiveness of the spectrogram enhancement algorithm, the structural similarity index (SSIM) method, which is a widely used metric in image analysis and quality assessment \cite{nilsson2020understanding} was used. It serves as a valuable tool for comparing two images by evaluating their structural similarity, taking into account luminance, contrast, and structure. SSIM quantifies how well the structural elements of an image, such as edges, textures, and patterns, are preserved when compared to a reference image. It produces a value between -1 and 1, where a score of 1 indicates a perfect similarity, meaning the two images are identical, while lower scores signify increasing dissimilarity. The SSIM index is calculated as follows:
\begin{equation}
	\text{SSIM}(x, y) = \frac{{(2\mu_x\mu_y + c_1)(2\sigma_{xy} + c_2)}}{{(\mu_x^2 + \mu_y^2 + c_1)(\sigma_x^2 + \sigma_y^2 + c_2)}}
\end{equation}
where $\mu_x$ is mean of $x$, $\mu_y$ is mean of $y$, $\sigma_x$ is standard deviation of $x$, $\sigma_y$ is standard deviation of $y$, $\sigma_{xy}$ is covariance between $x$ $y$. The constance $c_1$ and $c_2$ are defined by $c_i=(Lk_i)^2; i=1,2$, where $L$ is the dynamic range of pixel values, $k_1=0.01$ and $k_2= 0.03$ are default constants.
\subsubsection{Collection of Infrasonic Seismic Elephant rumbles}
A set of infrasonic seismic elephant rumbles are required initially to train a machine learning algorithm which can be used later for automated rumble detection. The geophone signal segments recorded while an elephant is rumbling (Fig.\ref{fig_rumble}) are used for this. Each of these signal segments is processed as described in previous sections to produce the spectrum ({\bf Algorithm 1}), which is then enhanced ({\bf Algorithm 2}).
\section {Identification of Elephant Rumbles in an Infrasonic Seismic Spectrogram}
A 20\% of the initial set of infrasonic seismic elephant rumble spectrograms is used as training data of which a set of features that characterizes the spectrogram is determined. This feature set is used as input to train machine learning. In this research, the following three methods were used to generate the spectrum features of the training set of spectrograms.
\begin{itemize}
	\item Mel-frequency cepstral coefficients
	\item Hjorth parameters
	\item Spectral energy distribution
\end{itemize}
\subsection{Mel-frequency Cepstral Coefficients}
The features in the spectrogram are identified by way of the Mel-frequency cepstral coefficients (MFCC)~\cite{m2014choice}. When extracting MFCCs, the signal is divided into overlapping frames of fixed 25 milliseconds. The frames are shifted by 50\% overlap to ensure continuity. On each frame, a  Hamming window is applied to taper the signal towards the frame boundaries. This is done to enhance the harmonics, smooth the edges, and reduce the edge effect while taking the discrete Fourier transform (DFT)~\cite{mutagi2004understanding} of the signal. Each windowed frame is converted into a magnitude spectrum by applying DFT as follows:
	\begin{equation}
		X(k) = \sum_{n=0}^{N-1} x(n) \cdot e^{-j\frac{2\pi}{N}nk}, \quad k = 0, 1, \ldots, N-1
	\end{equation}
where $X(k)$ is DFT of the input sequence $x(n)$, $N$ is number of samples in the sequence, and $k$ is frequency index.
The Mel spectrum is obtained by applying a Mel-filter bank to the Fourier-transformed signal. The approximation of the Mel frequency scale from the actual frequency is expressed as follows.
\begin{equation}
	m = 2595 \log_{10}\left(1 + \frac{f}{700}\right)
\end{equation}
where $m$ is Mel-frequency scale of the actual frequency $f$ in Hz. The Mel spectrum of the magnitude spectrum $X(k)$ is computed by multiplying the magnitude spectrum by each of the  triangular Mel weighting filter as follows.
	\begin{equation}
		s(m) = \sum_{k=0}^{N-1} |X(k)|^2 H_m(k) ; \quad 0 \leq m \leq M - 1
	\end{equation}
where $s(m)$ is filter bank output at index $m$, $X(k)$ is the DFT of the signal, $H_m(k)$ is the Mel filter bank coefficient at $k$ and $m$, $N$ is the number of DFT points, and $M$ is the number of Mel frequency bins. The triangular Mel weighting filters are defined as follows.
\begin{align}
H_m(k) =
	\begin{cases}
		0 & \text{if } k < f(m - 1) \\
		\frac{2(k - f(m - 1))}{f(m) - f(m - 1)} & \text{if } f(m - 1) \leq k \leq f(m) \\
		\frac{2(f(m + 1) - k)}{f(m + 1) - f(m)} & \text{if } f(m) < k \leq f(m + 1) \\
		0 & \text{if } k > f(m + 1)
	\end{cases}
\end{align}
	where $H_m(k)$ is the Mel filter bank coefficient at $k$ and $m$, $m$ is the Mel index, and $f(m)$ is the Mel frequency corresponding to index $m$.\par
	The discrete cosine transform (DCT) is applied to the log-compressed filterbank coefficients. Since most of the signal information is represented by the first few MFCC coefficients and the dataset is small, only the first few coefficients are extracted. Finally, the MFCC is calculated as follows.
	\begin{equation}
		c(n) = \sum_{m=0}^{M-1} \log_{10}(s(m)) \cos\left(\frac{\pi n (m - 0.5)}{M}\right)
	\end{equation}
	where $c(n)$ is cepstral coefficient at index $n$, $s(m)$ is filterbank output at index $m$, $M$ is the number of Mel frequency bins, and $C$ is the number of cepstral coefficients.
	\subsection{Hjorth Parameters}
	Hjorth parameters form a set of statistical measures used to characterize the time-domain properties of a signal\cite{wannawijit2019ecg}. From those parameters we have used Hjorth activity ($H_a$), Hjorth mobility ($H_m$), and Hjorth complexity ($H_c$) which are defined as follows.
	\begin{eqnarray}
		H_a &=& \frac{1}{N-1} \sum_{n=1}^{N}[x(n) - \mu]^2\\
		H_m &=& \sqrt{\frac{Var(x[n] - x[n-1])}{Var(x[n])}}\\
		H_c &=& \frac{H_m(x[n] - x[n-1])}{H_m(x[n])} 
	\end{eqnarray}
where $\mu$ is the mean of $x(n)$, and Var() is the variance operator.
\subsection{Spectral Energy Distribution}
To compute the spectral energy distribution, the spectrogram is split into frequency bands. In this research, twenty-five frequency bands were used. Let's denote the frequency band by $B_k;k=1,2,...25$. The energy within each frequency band $E_k$ is calculated as follows by summing the squared magnitudes of the spectrogram's elements that fall within $B_k$.
	\begin{equation}
		E_k = \sum_{i\in B_k} \sum_{j=0}^{N-1} |\text{spectrogram}[i, j]|^2
	\end{equation}
	where $i$ is index of frequency bin within $k$, and $j$ is index of time frame $(0,N-1)$.
\subsection{Machine Learning}
MFCC feature set can be used with the following machine learning algorithms.
\begin{itemize}
	\item{Ridge Classifier}
	\item{Support vector machine(SVM) - Linear}
	\item{Logistic Regression}
\end{itemize}
The Hjorth parameters can be used with the following machine learning algorithms
\begin{itemize}
	\item{Decision tree classifier}
	\item{AdaBoost classifier}
	\item{RandomForest classifier}
\end{itemize}
and the spectral energy distribution can be used with the following machine learning algorithms
\begin{itemize}
	\item{Light gradient boosting machine classifier (LGBM)}
	\item{Gradient boosting classifier}
	\item{AdaBoost classifier}
\end{itemize}
The open-source machine learning platform {\tt LasyPredict} was used to test the above machine learning algorithms on the corresponding feature sets. The remaining 80\% of the initial spectrograms were used as testing data. As shown in Fig. \ref{fig_ML} the {\tt LasyPredict} receives the feature set in MFCC, Hjorth, or SED format, and executes corresponding machine learning algorithms to produce performance in terms of accuracy, balance accuracy, and F1. Through this process, the best-performing combination of feature set and algorithm can be identified.
\begin{center}
\tikzstyle{startstop} = [rectangle, rounded corners, minimum width=3cm, minimum height=0.5cm, text centered, draw=black, fill=white!30]
\tikzstyle{process} = [rectangle, minimum width=3cm, minimum height=0.5cm, text centered, draw=black, fill=white!30]
\tikzstyle{decision} = [diamond, minimum width=3cm, minimum height=1cm, text centered, draw=black, fill=white!30]
\tikzstyle{arrow} = [thick,->,>=stealth]
	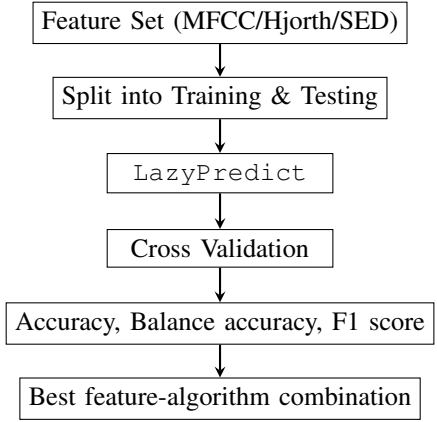
\begin{figure}
		\centering
	\begin{tikzpicture}[node distance=1cm]
		\node (step1) [process] {Feature Set (MFCC/Hjorth/SED)};
		\node (step2) [process, below of=step1] { Split into Training \& Testing};
		\node (step3) [process, below of=step2] {\tt LazyPredict};
		\node (step4) [process, below of=step3] { Cross Validation};
		\node (step5) [process, below of=step4] {Accuracy, Balance accuracy, F1 score};
		\node (step6) [process, below of=step5] {Best feature-algorithm combination};
		\draw [arrow] (step1) -- (step2);
		\draw [arrow] (step2) -- (step3);
		\draw [arrow] (step3) -- (step4);
		\draw [arrow] (step4) -- (step5);
		\draw [arrow] (step5) -- (step6);
	\end{tikzpicture}
\caption{Determination of the best feature-algorithm combination for machine learning using {\tt LazyPredict} platform}
\label{fig_ML}
\end{figure}
\end{center}
\section{Experimental Data Collection}
\begin{figure}[H]
	\centering   \includegraphics[width=1\columnwidth]{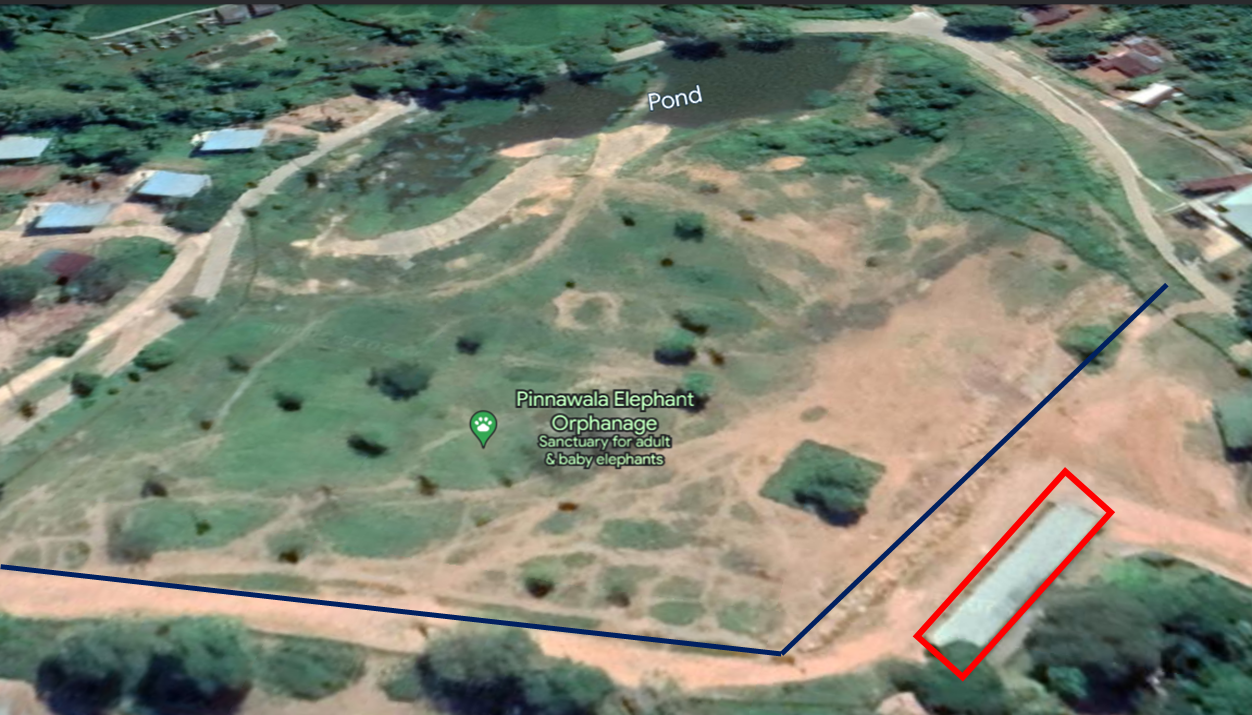}
	\caption{Free ranging area in Pinnawala elephant orphanage. The observation hall is the marked rectangle closer to the right-bottom corner}
	\label{fig:figure3nn}
\end{figure}

\begin{figure}
	\centering   \includegraphics[width=\columnwidth]{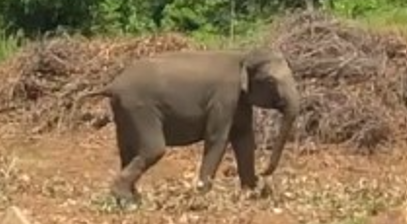}
	\caption{An elephant while rumbling - Pinnawala elephant orhpanage, Sri Lanka}
	\label{fig_rumble}
\end{figure}

\begin{figure*}[ht]
	\centering   \includegraphics[width=\columnwidth]{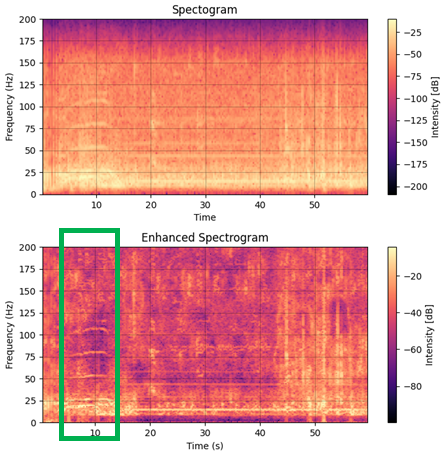}
	\includegraphics[width=\columnwidth]{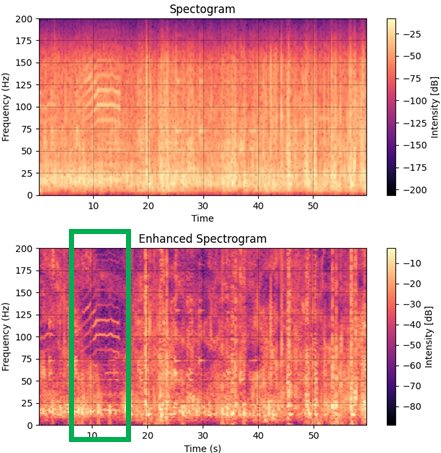}
	\caption{Spectrograms of two relatively strong rumbles verified by video-audio proof}
	\label{fig_Spectograms}
\end{figure*}

The experiment was conducted at the elephant orphanage, in Pinnawala, Sri Lanka (Fig. \ref{fig:figure3nn}), where there are about twenty-five elephants in a free-ranging area.
The elephants’ behavior was closely observed before the experiment. It was clear that in hot weather, elephants tend to be inactive and spend most of the day under trees. However, in normal weather, they become active and engaged. Seismic data was collected under average weather conditions using three equispaced geophone sensor nodes buried near the fence that is marked with straight lines in Fig.\ref{fig:figure3nn}. Elephant behavior was video recorded during the few-hour data capture. At the time the food truck arrived, one elephant rumbled quite noticeably as shown in Fig.\ref{fig_rumble} where it shows that the elephant stands literally on two legs exerting maximum pressure on the ground that would enhance the seismic rumble propagate through the ground.
The seismic data that is aligned with video and audio verification of rumbles were extracted from the continuous seismic signal record. The spectrogram clearly shows the frequency patterns of strong elephant rumbles as shown at the 10s mark of the top two illustrations if Fig.\ref{fig_Spectograms}.
\section{Results}
Altogether seven rumbles were detected within a few hours of data collection. Two of the rumbles ($\approx$20\%) were used as training data and the rest five rumbles ($\approx$80\%) were used as training data in the feature-algorithm evaluation process using {\tt lazyPredict} (Fig.\ref{fig_ML}) which produces the accuracy, balance accuracy and F1 score for each combination of feature extraction method and corresponding machine learning algorithm.
\subsection{Best Feature Extraction Method and Machine Learning Algorithm for Seismic Elephant Rumble Detection}
Table \ref{tab_MFCC} shows the rumble identification performance when the MFCC feature extraction method was used with the ridge classifier, SVM-linear, and Logistic regression as machine learning algorithms. 
\begin{table}[H]
	\centering
	\begin{tabular}{|c|c|c|c|}
		\hline
		\textbf{ML Algorithm} & \textbf{Accuracy} & \textbf{BA} & \textbf{F1 score} \\
		\hline
		Ridge Classifier & 97.05 ± 0.051 \% & 95.83\% & 95.65\% \\
		\hline
		SVM-linear & 94.04 ± 0.067\% & 90\% & 88.89\% \\
		\hline
		Logistic Regression & 93.85  ± 0.069\% & 84.16\% & 77.78\% \\
		\hline
	\end{tabular}
	\caption{Performance when MFCCs were used as features}
	\label{tab_MFCC}
\end{table}
Table \ref{tab_Hj} shows the rumble identification performance when Hjorth parameters were used as features and the Decision tree classifier, AdaBoost classifier, and RandomForest classifier were used as machine learning algorithms. 
\begin{table}[H]
	\centering
	\begin{tabular}{|c|c|c|c|}
		\hline
		\textbf{ML Algorithm} & \textbf{Accuracy} & \textbf{BA} & \textbf{F1 score} \\
		\hline
		Decision Tree Classifier & 93.84 ± 0.096\% & 83.33\% & 73.68\% \\
		\hline
		AdaBoost Classifier & 92.26 ± 0.097\% & 78.33\% & 74.04\% \\
		\hline
		RandomForest Classifier & 89.68  ± 0.056\% & 70\% & 57.14\% \\
		\hline
	\end{tabular}
	\caption{Hjorth parameters as features}
	\label{tab_Hj}
\end{table}
Table \ref{tab_SED} shows the rumble identification performance when spectral energy distribution was used as features and LGBM classifier, Gradient boosting classifier, and AdaBoost classifier were used as machine learning algorithms. 
\begin{table}[H]
	\centering
	\begin{tabular}{|c|c|c|c|}
		\hline
		\textbf{ML Algorithm} & \textbf{Accuracy} & \textbf{BA} & \textbf{F1 score} \\
		\hline
		LGBM Classifier & 86.67 ± 0.119\% & 64.17\% & 42.86\% \\
		\hline
		Gradient Boosting Classifier & 81.95 ± 0.095\% & 63.33\% & 40\% \\
		\hline
		AdaBoost Classifier & 80.24  ± 0.136\% & 58.33\% & 28.57\% \\
		\hline
	\end{tabular}
	\caption{Spectral energy distribution as features}
	\label{tab_SED}
\end{table}
The standard performance matrics are defined as follows.
\begin{eqnarray}
	Accuracy & =&  \frac{TP+TN}{TP+FP+TN+FN}\nonumber\\
	BA &=& \frac{Sensitivity+Specificity}{2}\nonumber\\
	F1~Score &=& \frac{2\times Precision\times Recall}{Precision + Recall}\nonumber
\end{eqnarray} in which
\begin{eqnarray}
	Sensitivity &=& \frac{TP}{TP+FN}\nonumber\\
	Specificity &=& \frac{TN}{TN+FP}\nonumber\\
	Precision &=&\frac{TP}{TP+FP}\nonumber\\
	Recall &=& \frac{TP}{TP+FN}\nonumber
\end{eqnarray}
where $TP$ is number of true positives, $FP$ is number of of false positives, $TN$ is number of true negatives, and $FN$ is number of false negatives. Based on the results in Table \ref{tab_MFCC}$\sim$Table\ref{tab_Hj}, MFCC with the Ridge classifier is the best combination for identifying elephant rumbles in seismic signals. The algorithms were trained using sci-kit learn 1.3.2 \cite{scikit-learn} Python library.
\subsection{Effectiveness of Spectrogram Enhancement}
One of the main contributions of this work is spectrogram enhancement presented in \ref{subsub-SE}. The degree of denoising in spectrogram enhancement is presented in Table \ref{tab_denoise}. 
\begin{table}[H]
	\centering
	\begin{tabular}{|c|p{3cm}|p{3cm}|}
		\hline
		\textbf{Spectrogram} & \textbf{SSIM with the initial spectrum ~\cite{zeppelzauer2015towards}} & \textbf{SSIM with the enhanced spectrum} \\
		\hline
		Fig.~\ref{fig_Spectograms}(left) & 0.86 &0.65\\
		\hline
		Fig.~\ref{fig_Spectograms}(right)& 0.89 & 0.66 \\
		\hline
	\end{tabular}
	\caption{Denoising effect in the spectrogram enhancement process, measured in SSIM score}
	\label{tab_denoise}
\end{table}
A low SSIM score indicates that fewer structural details are retained after spectrogram enhancement. The SSIM value has dropped by about 20\% with the spectrum enhancement.
\section{Conclusion}
The electronic circuitry and signal conditioning techniques have been developed quite successfully leading to acquiring seismic elephant rumbles in good quality. A few elephant rumbles were visually and audibly noticed during the data collection. Their seismic infrasonic copies were extracted from the geophone data, and the spectrograms were determined. The spectrogram enhancement has reduced noise in the seismic data quite significantly leading to achieving higher accuracy in feature detection and machine learning. Using {\tt LazyPredict}, it was found that the MFCC feature extraction method and Ridge classifier ML algorithm give the best performance in automatic elephant rumble detection
.\par
The proposed elephant rumble identification method could be used to devise an early warning system for the villages troubled by wild elephants. The circuits involved are low-power and low-cost. It could be made even more cost-effective and less power-consuming by replacing the Raspberry Pi board with a Raspberry Pi zero board, which extends the battery life in field deployment. For the practical deployment, the best ML algorithm, Ridge classifier can be implemented on an edge device using {\tt scikit-learn} library in Python. It is important to mention that elephants do not rumble often, hence rumble detection will not lead to the development of a perfect early warning system. The elephant localization using seismic rumbles is yet to be investigated, though some early work has been reported.
\section*{Acknowledgment}
The authors would like to thank Dr. Daniela Hedwig, Director of the Elephant Listening Project, Dr. Geoffrey Abers, Chair of the Earth and Atmospheric Sciences, both at Cornell University, USA, and Dr. Chase A. LaDue, at Oklahoma City Zoo and Botanical Garden, USA for their insightful guidance.\par
Dr. Lori Lenard, Dr. Ronnie Coffman, and Dr. Richard Cahoon are acknowledged for facilitation and guidance provided for this research during the Fulbright Fellowship of the principal investigator Dr. Rohan Munasinghe at the Dept of Global Development, CALS, Cornell University, Ithaca, NY 14853, USA.\par
The authors would also acknowledge the staff of the National Zoological Department, and the Elephant Orphanage, Pinnawala, Sri Lanka for their approval and invaluable support for the data collection.


%





\ifCLASSOPTIONcaptionsoff
\newpage
\fi






\bibliographystyle{IEEEtran}
\bibliography{rumble}
%






\vfill
\balance
\end{document}